\documentclass{elsart5p}%
\usepackage{graphics}
\usepackage{graphicx}
\usepackage{amssymb}%
\usepackage{amsmath}%
\usepackage{amsfonts}
\begin{document}
\begin{frontmatter}
%
%
\title{2D-MIT as self-doping of a Wigner-Mott insulator}
\author[cor1]{S. Pankov}
\author{V. Dobrosavljevic}
\corauth[cor1]{Corresponding author; email: pankov@magnet.fsu.edu}
\address{National High Magnetic Field Laboratory, Florida State
University, Tallahassee, FL 32306}

\begin{abstract}
We consider an interaction-driven scenario for the two-dimensional
metal-insulator transition in zero magnetic field (2D-MIT), based
on melting the Wigner crystal through vacancy-interstitial pair
formation. We show that the transition from the Wigner-Mott
insulator to a heavy Fermi liquid emerges as an instability to
self-doping, resembling conceptually the solid to normal liquid
transition in He3. The resulting physical picture naturally
explains many puzzling features of the 2D-MIT.
\end{abstract}
\end{frontmatter}

A series of fascinating experiments, as first performed some ten
years ago by Kravchenko and co-workers~\cite{kravchenko95}, have
deeply changed our thinking about the two dimensional electron gas
(2DEG).  These and many later experiments demonstrated
convincingly that the 2DEG can exhibit typical metallic
behavior~\cite{kravchenko04} above a well defined critical density
$n_c$. One of the most prominent features of this metallic phase
is an unprecedented resistivity drop, which is found only in the
low density regime $n\gtrsim n_c$ close to the transition. These
findings suggested that a well-defined metal-insulator transition
(MIT) may exist even in two dimensions, in contrast to long
held-beliefs based on the theories for noninteracting disordered
electrons.

These experiments are typically performed at such low electron
density where the relative strength of the Coulomb interaction is
so large ($r_s \gtrsim 10$), that the localization processes could
conceivably be suppressed by interaction effects. A diffusion-mode
theory describing such interaction renormalizations at weak
disorder has been developed by Finkelshtein and
Punnoose~\cite{punnoose02}, suggesting that sufficiently strong
interactions may indeed stabilize the metallic phase. However,
this theory can provide guidance only within a narrow diffusive
regime restricted to very low temperatures.

In contrast, the most striking experimental results have been
established in a broad parameter range well outside this regime,
and are most pronounced {\em in the cleanest
samples}~\cite{kravchenko04}. In particular, the {\em best}
established experimental signature of the transition relies on
careful effective mass measurements, which is found to diverges as
$m^*\propto(n-n_c)^{-1}$, while the Lande $g^*$-factor remains
largely unrenormalized. Such phenomena cannot be understood within
a low-energy diffusion mode theory, which relies on Anderson
localization to produce an insulating phase.

\begin{figure}
\vspace{.31cm}
\includegraphics[width=2.9in]{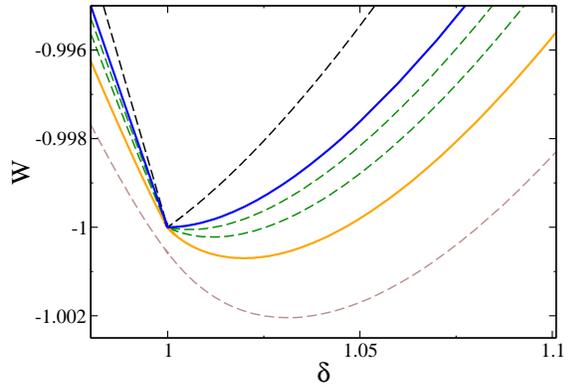}
\vspace{.0cm} \caption{Evolution of the free energy profile
$W[\delta]$, as the electron density is increased across the MIT
(from top to bottom). The cusp corresponds to the Wigner solid.
The self-doped transition (blue line) takes place before the
instability emerges at half-filling ($\delta=0$, orange line) and
the Wigner insulating state is replaced by a heavy Fermi liquid.}
\end{figure}

A fundamental question is thus posed by these experiments: what is
the basic mechanism that drives the metal-insulator transition in
these systems? Does one have to rely on disorder effects at all in
the zero-th order approximation, or can one understand the most
important experimental features by interaction effects alone? In
this work, we concentrate solely on the effects of strong Coulomb
interactions in the clean limit, and try to establish which
experimental features can be explained by entirely ignoring the
disorder.

We approach the transition from the insulating side, starting with
the Wigner-Mott insulator, and examine its melting by quantum
fluctuations as density increases. These are believed to be
dominated~\cite{tanatar89} by the vacancy-interstitial pair
excitations on top of the classical triangular lattice
configuration.  Within the Wigner crystal, the electrons are
tightly bound, and the vacancy-interstitial excitations can be
well represented by a mapping to a two-band Hubbard (i.e.
charge-transfer) model, respectively corresponding to the lattice
and the interstitial electrons.  As the density increases, the
charge-transfer gap eventually closes, and a metal-insulator
transition assumes the character of a Mott-metal-insulator
transition, leading to a strongly correlated metallic state on the
metallic side.

Our model Hamiltonian reads:

\begin{multline}
H=\sum_{i\sigma}e_f f_{i\sigma}^{\dagger}f_{i\sigma}
+e_c c_{i\sigma}^{\dagger}c_{i\sigma}
-\sum_{ij\sigma}t_{ij}c_{i\sigma}^{\dagger}c_{j\sigma}\\
+\sum_{i\sigma}V(f_{i\sigma}^{\dagger}c_{i\sigma}
+c_{i\sigma}^{\dagger}f_{i\sigma})
+\sum_{i}Uf_{i\uparrow}^{\dagger}f_{i\uparrow}
f_{i\downarrow}^{\dagger}f_{i\downarrow}
\label{hamiltonian}
\end{multline}
where $f^{\dagger}$, $f$ and $c^{\dagger}$, $c$ are creation and
annihilation operators for site and interstitial electrons
respectively. It is assumed that the inter-cell hopping $t_{ij}$
exists only between interstitial orbitals, while only the site
electrons are subject to onsite Coulomb repulsion $U$, and the
interstitial orbitals are coupled to the site orbitals via
hybridization $V$. The local electrostatic potentials for the two
bands are denoted by $e_f$ and $e_c$.

We model the strong onsite repulsion by exclusion of double
ocupancy, using the standard slave-boson mean-field formalism. The
free energy per electron then reads:

\begin{multline}
W[\lambda,Z,\mu,\delta]=\\
-\frac{2T}{1-\delta}\sum_{lk}\ln{(1+\exp{(-(\tilde\varepsilon_{lk}-\mu)/T)})}\\
+\frac{\lambda}{1-\delta}(Z-1)+\mu \label{freeenergy}
\end{multline}
where $\varepsilon_{lk}$ are renormalized band energies, $\lambda$
is the Lagrange multiplier enforcing the slave-boson constraint,
$Z$ is the quasiparticle weight, $\mu$ is the chemical potential.
The chemical potential $\mu$ is an internal parameter here,
arising similarly to $\lambda$, from constraining the electron
density per unit area. The self-doping $\delta$ measures deviation
in the number of electrons (per elementary cell) from the half
filling. In the classical limit of low electron density
$\delta=0$, but it becomes $\delta\ne0$ at higher density, where
quantum effects are important. This is reminiscent of the liquid
solid transition in He3 \cite{vollhardt84}. The equations of the
state are given by the saddle point of the free energy
$W[\lambda,Z,\mu,\delta]$.

A peculiarity of this model is that the band energies are not
fixed, but are self-consistently determined through their
dependence on the occupation of the site and interstitial
orbitals. It immediately follows that the system is unstable to
the self-doping at the MIT. Indeed, by carefully accounting for
the electrostatic energy balance due to such charge transfer, we
find that the free energy takes a lower value (relative to the
classical value at $\delta=0$) on one of the branches at
$\delta\ne0$ {\em before} the transition at half filling is found
(see Fig. 1).

With appropriately chosen model parameters we can explain the
basic experimental results,  which otherwise cannot be captured by
the diffusion mode theory. Similarly to what is seen
experimentally, we observe strong renormalization of the effective
mass near the transition $m^* \sim (n-n_c)^{-1}$. In this strongly
correlated regime any extrinsic disorder is very effectively
screened by interaction effects~\cite{screening}, providing a
plausible scenario for the large resistivity drop~\cite{resdrop}.
The magneto-resistance data are also naturally explained by our
two band model. The experiment \cite{jaroszynski04} has shown that
in high parallel magnetic fields the transport takes place by
activated processes, with an activation gap that vanishes linearly
at some density $n_1>n_c$. In the strong field our model reduces
to a trivial model of noninteracting spinless electrons. For the
density $n<n_1$ the system is a band insulator with the lower band
completely filled. The bands broaden as the density increases, and
the insulating gap closes linearly at some density $n_1 > n_c$.

Our static lattice model does not capture the collective charge
density fluctuations -- the phonons of the Wigner crystal. Because
the Coulomb interaction is long ranged, these collective modes are
very soft, and play an important role in renormalizing the model
parameters. It is this strong renormalization of the charge
transfer gap $e_c-e_f$ that pushes the transition to such low
density~\cite{lenac95} ($r_s \gg 1$)), in a fashion that is
conceptually very similar to the formation of the Coulomb
gap~\cite{pankov}. At present, these effects are incorporated in
through the choice of the electrostatic parameters of our model.
In future work, we would like to systematically incorporate these
soft collective modes, the effects of which are currently included
in a semi-phenomenological fashion. This program can be achieved
by a variety of methods, including extended dynamical mean field
approaches (EDMFT), by exploiting unique properties of the Coulomb
potential, and by employing the techniques recently developed in
the Coulomb glass context~\cite{pankov}.


\begin{thebibliography}{10}                                                                                                %

\bibitem{kravchenko95}
S. V. Kravchenko et al., Phys. Rev. B \textbf{51}, 7038 (1995).

\bibitem{kravchenko04}
For a review, see S. V. Kravchenko and M. P. Sarachik, Rep. Prog.
Phys. \textbf{67}, 1 (2004).

\bibitem{punnoose02}
A. Punnoose and A. M. Finkelstein, Phys. Rev. Lett. \textbf{88},
016802 (2002).

\bibitem{tanatar89}
B. Tanatar and D. M. Ceperley Phys. Rev. B \textbf{39}, 5005
(1989).

\bibitem{vollhardt84}
D. Vollhardt, Rev. Mod. Phys. \textbf{56}, 99 (1984).

\bibitem{lenac95}
Z. Lenac and   M. Sunjic, Phys. Rev. B {\bf 52}, 11238 (1995).

\bibitem{screening} D. Tanaskovi\'c, et al.,
Phys. Rev. Lett. \textbf{91}, 066603  (2003).

\bibitem{resdrop}M. C. O.
Aguiar et al., Europhys. Lett. \textbf{ 67}, 226 (2004).


\bibitem{jaroszynski04}
J. Jaroszynski et al, Phys. Rev. Lett. \textbf{92}, 226403 (2004).

\bibitem{pankov} S. Pankov and V. Dobrosavljevi\'c, Phys. Rev. Lett. {\bf 94}, 046402
(2005).

\end{thebibliography}
\end{document}